\documentclass[twocolumn,showpacs,superscriptaddress,aps,prl,floatfix]{revtex4-1}
\usepackage{graphicx}
\usepackage{amsmath,amssymb}
\usepackage[utf8]{inputenc}
\usepackage[T1]{fontenc}
\usepackage{lmodern}
\usepackage{natbib}
\usepackage{hyperref}
\usepackage{color}
\hypersetup{colorlinks=true,citecolor={blue},linkcolor={blue},urlcolor={blue}}
\usepackage{units}
\usepackage[english]{babel}
\usepackage{dcolumn}
\usepackage{bm}
\usepackage{xfrac}
\usepackage[normalem]{ulem}
\usepackage{nicefrac}
\usepackage{bm}
\usepackage{soul}
\setstcolor{red}
\usepackage{xcolor,cancel}
\newcommand{\Fig}[1]{Fig.\,\ref{#1}}

\begin{document}
\renewcommand{\thefootnote}{\alph{footnote}}	

\title{Coherence Revivals of a Spinor Polariton Condensate from Self-induced Larmor Precession}
\date{\today}

\author{A. Askitopoulos}\email{A.Askitopoulos@skoltech.ru}
\affiliation{Skolkovo Institute of Science and Technology, Moscow, Bolshoy Boulevard 30, bld. 1, 121205, Russia}
\author{H. Sigurdsson}\email{H.Sigurdsson@soton.ac.uk}
\affiliation{Skolkovo Institute of Science and Technology, Moscow, Bolshoy Boulevard 30, bld. 1, 121205, Russia}
\affiliation{School of Physics and Astronomy, University of Southampton, Southampton, SO171BJ, United Kingdom}
\author{I. Gnusov}
\affiliation{Skolkovo Institute of Science and Technology, Moscow, Bolshoy Boulevard 30, bld. 1, 121205, Russia}
\author{S. Alyatkin}
\affiliation{Skolkovo Institute of Science and Technology, Moscow, Bolshoy Boulevard 30, bld. 1, 121205, Russia}
\author{L. Pickup}
\affiliation{School of Physics and Astronomy, University of Southampton, Southampton, SO171BJ, United Kingdom}
\author{N.A. Gippius}
\affiliation{Skolkovo Institute of Science and Technology, Moscow, Bolshoy Boulevard 30, bld. 1, 121205, Russia}
\author{P.G. Lagoudakis}
\affiliation{Skolkovo Institute of Science and Technology, Moscow, Bolshoy Boulevard 30, bld. 1, 121205, Russia}
\affiliation{School of Physics and Astronomy, University of Southampton, Southampton, SO171BJ, United Kingdom}
\begin{abstract}
First order coherence measurements of a polariton condensate, reveal a regime where the condensate pseudo-spin precesses persistently within the driving optical pulse. Within a single 20 $\mu$s optical pulse the condensate pseudo-spin performs over $10^5$ precessions with striking frequency stability. The condensate maintains its phase coherence even after a complete precession of the spin vector, making the observed state by a definition a spin coherent state. The emergence of the precession is traced to the polariton interactions that give rise to a self-induced out-of-plane magnetic field that in turn drives the spin dynamics. We find that the Larmor oscillation frequency scales with the condensate density, enabling external tuning of this effect by optical means. The stability of the system allows for the realization of integrated optical magnetometry devices with the use of materials with enhanced exciton $g$-factor and can facilitate spin squeezing effects and active coherent control on the Bloch sphere in polariton condensates.
\end{abstract}

\maketitle
The collapse and revival of a matter-wave field, observed in lattices of atomic Bose-Einstein condensates (BECs), has been suggested as a utilitarian process for the creation of highly entangled states~\cite{greiner_collapse_2002}, and revealed the influence of the underlying granulated quantum nature to the dynamics of the system.  The revival of coherence in these systems was anticipated due to Bloch oscillations in an optical lattice~\cite{ben_dahan_bloch_1996}, while latter studies showcased that the system can undergo tens of thousands of these oscillations within its coherence time~\cite{gustavsson_control_2008}, enabling the study of many body transport~\cite{haller_inducing_2010}, many body strongly correlated quantum phases~\cite{will_time-resolved_2010} and quantum-chaos dynamics~\cite{meinert_interaction-induced_2014}. Although the demonstration of the BEC phase transition in polaritons~\cite{kasprzak_bose-einstein_2006, balili_bose-einstein_2007}, photons~\cite{klaers_boseeinstein_2010} and plasmons~\cite{hakala_boseeinstein_2018}, has facilitated the study of these effects in different material platforms, the collapse and revival of the matter wave field in these systems in the non-linear regime has remained elusive. 

Exciton-polaritons (here-forth polaritons) are two-component bosonic quasi-particles that can condense into a macroscopically occupied state~\cite{kasprzak_bose-einstein_2006,balili_bose-einstein_2007}. The system order parameter is related to the emergence of a well defined pseudo-spin state~\cite{kavokin_quantum_2004}, which along with the strong interparticle interaction, has enabled the observation of spin switching and hysteresis regimes under quasi-resonant excitation~\cite{amo_excitonpolariton_2010, paraiso_multistability_2010}. From the perspective of manipulating a coherent many-body state, non-resonant injection of polaritons results in some complications due to the order parameter depolarizing and dephasing from interactions of the condensate with the background reservoir of non-condensed particles~\cite{askitopoulos_giant_2019, gnusov_optical_2020}. However, the development of all-optical trapping techniques, where reservoir and condensate become spatially separated~\cite{askitopoulos_polariton_2013, cristofolini_optical_2013, dall_creation_2014, askitopoulos_robust_2015}, enabled better harnessing of the condensate coherence properties and spin-degrees of freedom. Today, optical trapping has revealed intriguing phenomena like spin switching and inversion, spin bistability and spin bifurcations~\cite{ohadi_spontaneous_2015, askitopoulos_nonresonant_2016, pickup_optical_2018, del_valle-inclan_redondo_observation_2019} all under non-resonant excitation. This allowed for the proposition of polariton condensates for a number of novel spinoptronic devices, such as optoelectronic spin switches and spin valves~\cite{dreismann_sub-femtojoule_2016, askitopoulos_all-optical_2018}. 

Moreover, the condensate spin dynamics under pulsed optical excitation, have revealed the existence of spin quantum beats~\cite{renucci_microcavity_2005}, as well as dynamic spin precession due to the self-induced Larmor effect~\cite{solnyshkov_nonlinear_2007,laussy_effects_2006, demenev_polarization_2017}, arising from the interaction induced effective, out-of plane, magnetic field. The out-of-equilibrium nature of the system in this pulsed (transient) regime leads to a time dependent modification of the system non-linearity, and thus self-induced field. Due to the ultra-fast polariton decay rate ($\approx\unit[5-30]{ps}$ particle lifetime), the precession is quickly dampened. However, under dynamic equilibrium, when the condensate losses are continuously balanced by an external optical pump, these precession dynamics can be expected to survive much longer but have, so far, not been directly observed. Theoretical investigations~\cite{li_incoherent_2015} as well as recent spin noise experiments~\cite{ryzhov_spin_2020} indicate that a persistent precession of the condensate pseudo-spin can indeed take place which would further underline the potential applicability of the system in different fields ranging from magnetometry to coherent control on the Bloch sphere. 
\begin{figure*}[t]
	\center	
	\includegraphics[width=2.0\columnwidth]{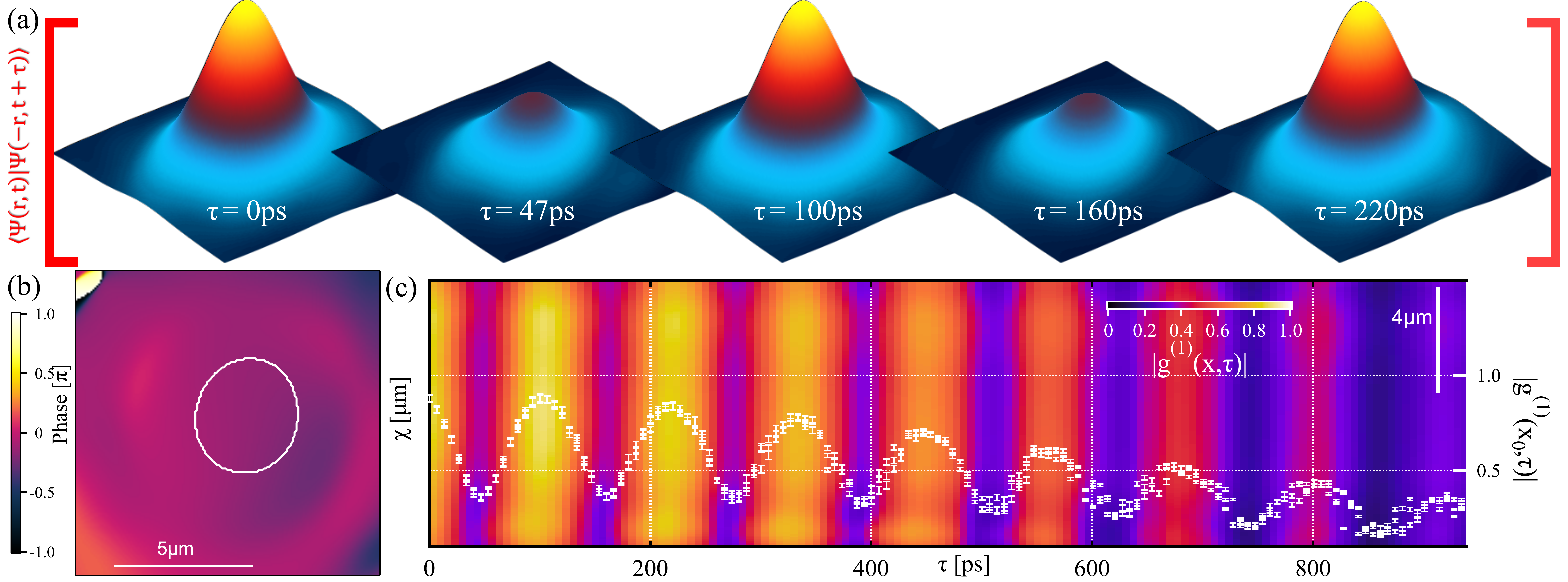}
	\centering
	\vspace{-10pt}
	\caption{(a) Condensate $|\langle \Psi^\dagger(\mathbf{r}, t) \Psi(-\mathbf{r},t + \tau) \rangle|$ reconstructed through Fourier analysis for varying $\tau $. (b) Condensate spatial phase difference $\Delta \Phi(\mathbf{r},\tau)$ for $\tau=0$ showing a uniform phase across the whole condensate. White line indicates the condensate FWHM. (c) First order coherence function $|g^{(1)}(x, \tau)|$ versus $\tau$ and $x$ for a central slice of the condensate. White dots are the averaged  $|g^{(1)}(\tau)|$ within the FWHM region of (b) plotted versus the right axis and error bars are the standard deviation.}
	\label{fig1}
	\vspace{-10pt}
\end{figure*}

In this letter, we report the observation of partial collapse and revival of coherence in a polariton condensate in the dynamic equilibrium regime where gain and dissipation are continuously balanced. The effect of coherence revivals is traced to the self-induced Larmor precession of the condensate spin. This is evidenced further by tuning the population of the condensate which controls the spin precession rate. The spin precession persists with prodigious stability throughout the condensate lifetime. Condensate lifetime here denotes the duration that the condensate is excited and in dynamic equilibrium, which coincides with the excitation pulse duration of $\unit[20]{\mu s}$. Despite the effects of gain, dissipation and dephasing, the condensate performs more than $10^5$ cycles in every experimental realization. As the condensate coherence time is considerably longer than the spin precession period, the system can de-facto be described as a spin coherent state that has been suggested as a basis for polariton condensates in quantum information applications~\cite{byrnes_macroscopic_2012, Cuevas_SciAdv2018, ghosh_quantum_2020}. Despite, the non-Hermitian nature of the polariton Hamiltonian, this does not preclude the existence of PT-symmetry as has already been discovered in other non-Hermitian photonic systems~\cite{lumer_nonlinearly_2013, el-ganainy_non-hermitian_2018}.



Using a single mode, ultra-narrow linewidth, continuous wave (CW) circularly polarized coherent optical source, we create an optical annular harmonic trap~\cite{askitopoulos_polariton_2013} on a semiconductor microcavity sample~\cite{cilibrizzi_polariton_2014} and non-resonantly (detuned $\Delta E= \unit[110]{meV}$ from the lower polariton mode) excite carriers into the system. Previous studies have revealed that in this configuration above condensation threshold the system exhibits a coherence time in the order of nanoseconds~\cite{askitopoulos_giant_2019}. This allows one to use first order correlation measurements to study complex dynamics of the condensate over much longer time scales. The circular polarization of the pump induces a spin imbalance in the reservoir resulting in a dominantly co-polarized condensate above threshold~\cite{ohadi_spontaneous_2015, gnusov_optical_2020}. Our sample region corresponds to exciton fraction $|X|^2=0.35$, at a photon-exciton detuning $\Delta=\unit[-1.75]{meV}$. With the use of a Michelson interferometer in a retro-reflector configuration, we perform single shot time delayed interferometry. The optical excitation is time shaped into 20 $\mu$s square pulses with an acousto-optic modulator and for every condensate instance we record a single interferogram with a 5 $\mu$s integration time. Through Fourier analysis (see Supplementary Information (SI)) of the interference fringes we then extract the condensate spatial phase difference map $\Delta\Phi(\mathbf{r,\tau})=\Phi(\mathbf{r},t)-\Phi(-\mathbf{r},t+\tau)$ where $\Phi$ is the weighted sum of the phase of the two spin components, and first order correlation function ,
\begin{equation}
g^{(1)}(\mathbf{r}_1,  \mathbf{r}_2, \tau) = \frac{\langle \Psi^\dagger(\mathbf{r}_1, t) \Psi(\mathbf{r}_2,t +\tau) \rangle}{\sqrt{\langle |\Psi(\mathbf{r}_1,t)|^2 \rangle \langle |\Psi(\mathbf{r}_2,t +\tau)|^2 \rangle}},
\end{equation}
where $\Psi = (\psi_+,\psi_-)^t$ is the condensate spinor and $\langle . \rangle = \frac{1}{T} \int_0^T dt$ denotes time average over a single realization of the condensate, shown in~\Fig{fig1}. 

Scanning the time delay $\tau$ we observe regular periodic oscillations of the $|\langle \Psi^\dagger(\mathbf{r}, t) \Psi(-\mathbf{r},t + \tau) \rangle|$ amplitude in Fig.~\ref{fig1}(a) with a periodicity of approximately \unit[110]{ps}. The phase map of the condensate shows a flat distribution within the full width half maximum (FWHM) region of the condensate for all time delays (see Supplementary video 1, \hyperref{https://www.dropbox.com/s/isaggsoiijxn229/SV1_g1extractionHR.avi?dl=0}{SV1}{SV1}{SV1}), while the extracted first order coherence also displays similar oscillations shown in~\Fig{fig1}(b,c). Notably, the amplitude of these coherence revivals is almost half the coherence amplitude, indicating that the dynamic beating we observe is between non-orthogonal states. The extended coherence of our system in this regime enables us to observe up to 8 revivals of the coherence for the delay range available in our configuration. However, the fact that we are able to observe these beatings although they are considerably faster (by a factor of $> 10^5 $) than the condensate lifetime evidences the frequency stability and perseverance of these oscillations within the condensate lifetime. Indeed the only available effect that can compromise the stability of the system are optical heating effects, which will slowly but steadily change the lower polariton energy level and polariton interaction strength. We point out that spin hysteresis effects, in the same configuration, governed by the same underlying mechanisms and limitations, have demonstrated their perseverance even up to 100's of milliseconds~\cite{pickup_optical_2018}. 

\begin{figure}[t!]
	\center	
	\includegraphics[width=1.0\columnwidth]{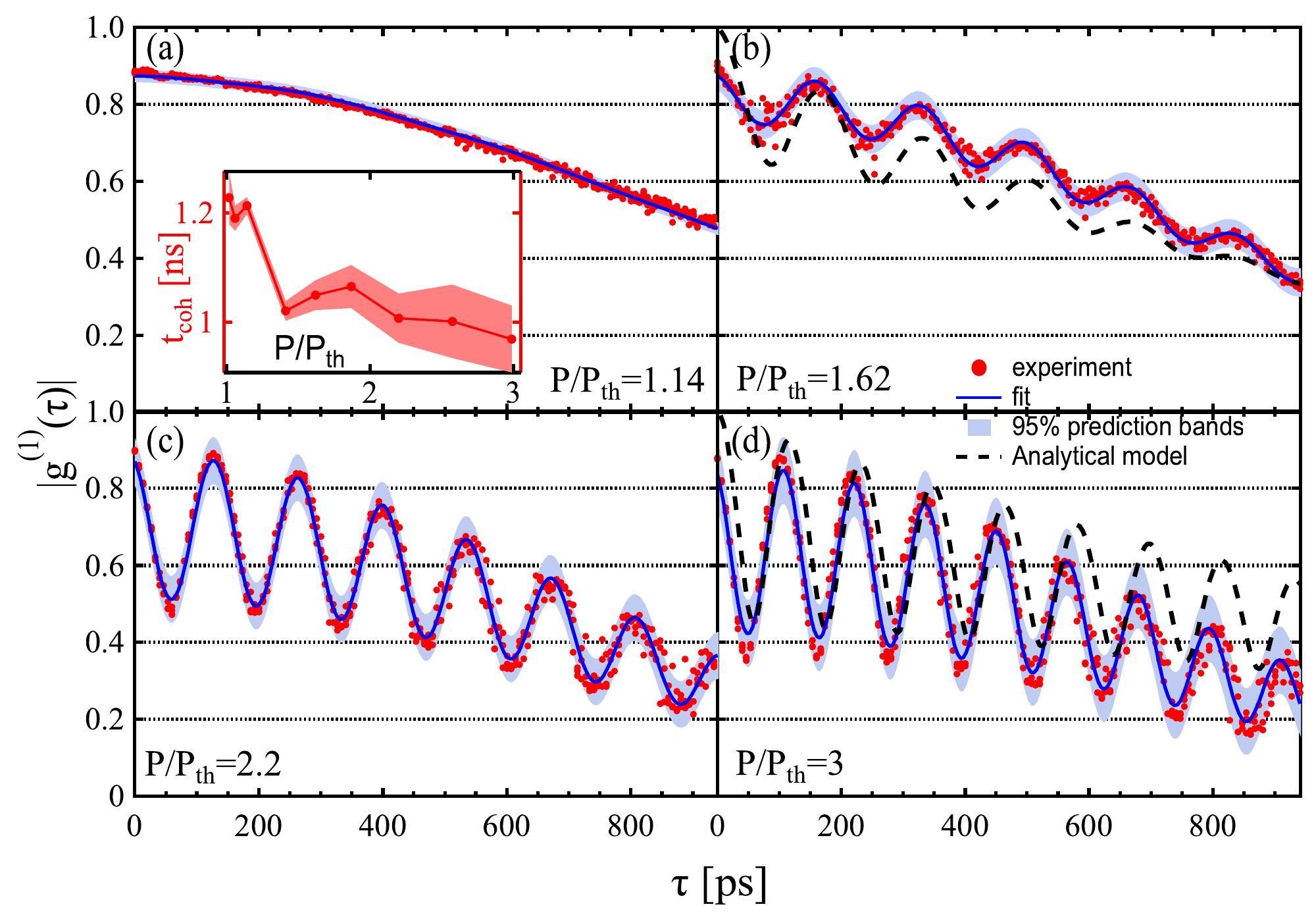}
	\centering
	\vspace{-24pt}
	\caption{$|g^{(1)}(\tau)|$ for increasing condensate density, $P=1.14P_{th}$ (a), $P=1.62P_{th}$ (b), $P=2.2P_{th}$ (c), $P=3P_{th}$ (d). Red circles correspond to $|g^{(1)}(\tau)|$ averaged within the condensate FWHM region, continuous blue line is the fit and shaded area the 95\% prediction band of the fitting parameters.  Dashed black lines in (b) and (d) are the results from the simulation described in the text. Inset in (a) shows the extracted coherence time $t_{coh}$ for increasing density. }
	\vspace{-15pt}
	\label{fig2}
\end{figure}
Performing the same experiment, but for a condensate density just above threshold ($P=1.14P_{th}$), we observe the absence of coherence revivals~\Fig{fig2}(a). As this indicates that the condensate density is a decisive factor in the emergence of the coherence revivals, we gradually increase it, thus raising the total interactions in the system. We subsequently observe the emergence of the periodic oscillations of the first order coherence function that, with increasing condensate particle number, become faster and more pronounced~\Fig{fig2}(b-d). To extract the relevant parameters of the coherence revivals, we approximate the first order coherence as $|g^{(1)}(\mathbf{r},\tau) |= g^{(1)}_{env}(\mathbf{r},\tau) (1- A\sin^2\left(\omega \tau\right))$, where $\omega$ and $A$ are fitting parameters and $g^{(1)}_{env}(\mathbf{r},\tau)$ is an analytic coherence function that takes into account particle number fluctuations and the inter-particle interaction strength causing the decay of coherence, described in~\cite{whittaker_coherence_2009, askitopoulos_giant_2019}. In order to limit the number of free parameters, we assume $g^{(1)}_{env}(\tau)$ as a Gaussian that incorporates all such decoherence effects into a single parameter $t_{coh}$ describing the condensate coherence time (inset in~\Fig{fig2}(a))~\cite{whittaker_coherence_2009} and fit the experimental $|g^{(1)}(\tau)|$ in~\Fig{fig2}(a-d). We note that the observed revivals are well approximated with this single beat frequency $\omega$ even for relatively low oscillation amplitude. This points to the presence of a coherent dual mode (two color) condensate, that is stable and present throughout its lifetime.

\begin{figure}[h!]
	\center	
	\includegraphics[width=1.0\columnwidth]{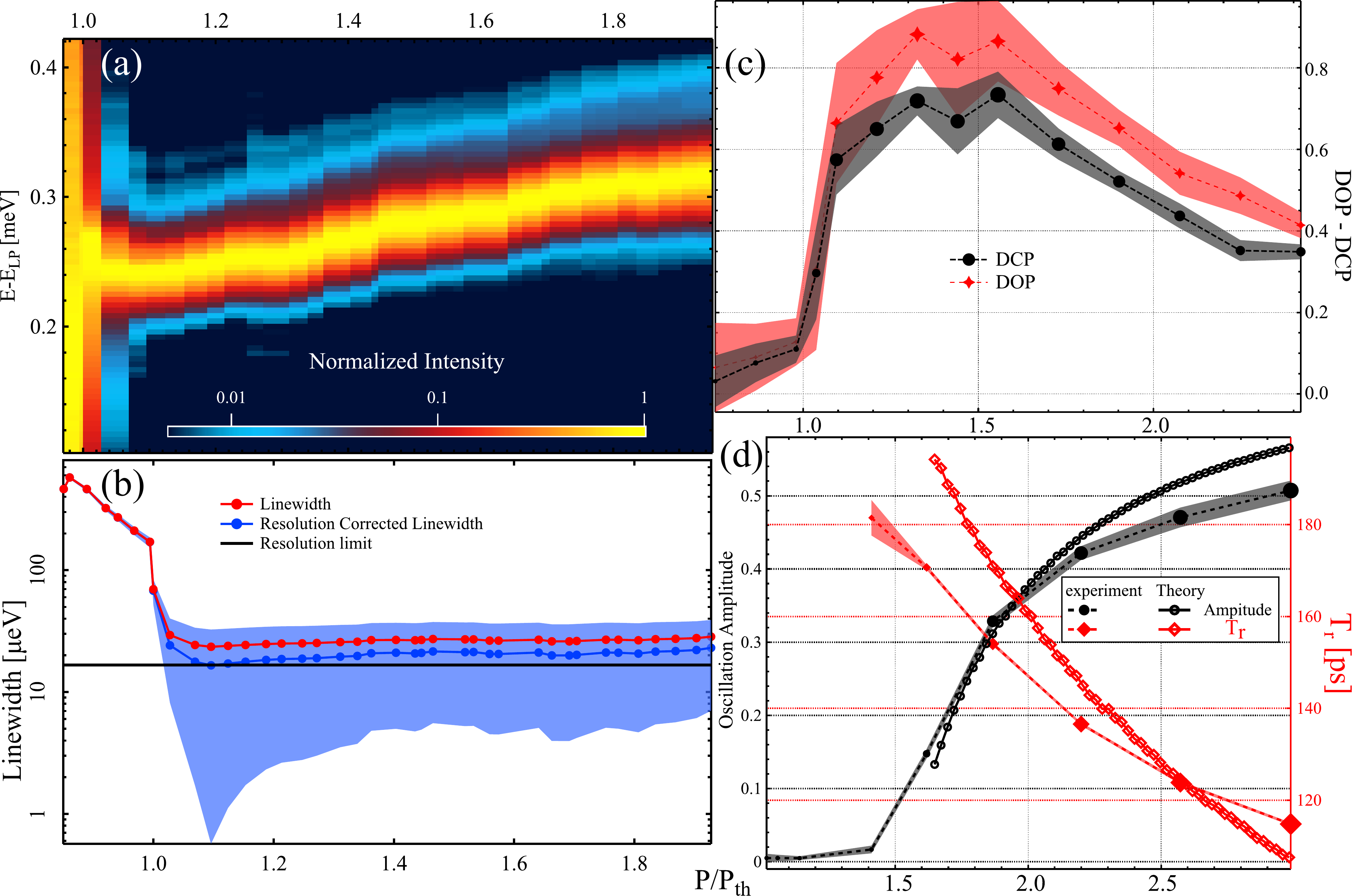}
	\centering
	\vspace{-15pt}
	\caption{(a) Condensate energy vs excitation power in units of threshold power ($P_{th}$), in a logarithmic colorscale. The spectrum for each excitation power is extracted from energy dispersion imaging at zero in-plane momentum and have been intensity normalized for clarity. Condensate energy is measured relative to the zero momentum free polariton state. (b) Corresponding extracted linewidth (red dots) also taking into account resolution correction (blue dots), blue shaded region is the error fit. (c) DOP and DCP of the condensate vs excitation power. (d) Extracted coherence revival parameters from the fit of $|g^{(1)}(\tau)|$ and corresponding parameters from simulation (filled and open symbols respectively). Oscillation amplitude (black dots, left axis), period (red diamonds, right red axis).}
	\label{fig3}
\end{figure}

We do not observe any spatial reshaping effects in $|\langle \Psi^\dagger(\mathbf{r}, t) \Psi(-\mathbf{r},t + \tau) \rangle|$ that would be indicative of mixing of the ground state with higher order modes of the potential. This is further corroborated by the non-orthogonality of the beating modes, as previously mentioned. Nevertheless, we perform an additional spectroscopic study of the condensate energy and linewidth. In~\Fig{fig3}(a,b) we show that even far above threshold, the potential contains a single energy mode with a resolution limited linewidth. To further verify that the observed coherence revivals originate from the precession of the condensate pseudo-spin, we record the evolution of the time averaged degree of circular polarization (DCP) and degree of polarization (DOP) of our system. Just above threshold we observe a high degree of circular polarization~\Fig{fig3}(c), which gradually declines after $1.5P_{th}$. The condensate DOP also starts to degrade for the same excitation power, indicating that the time averaged pseudo-spin of the condensate $\langle\mathbf{S} \rangle$ (see Eq.~\eqref{eq.S}) is not shifting towards a fixed attractor in the equatorial plane of the Bloch sphere (which would explain the drop in DCP but not DOP). This effective depolarization of the system, perfectly coincides with the emergence of the revivals of coherence. 
The extracted period and amplitude of the $|g^{(1)}(\tau)|$ oscillations are displayed in~\Fig{fig3}(d) for varying condensate density. Although the oscillation amplitude sharply increases past the critical value of $1.5P_{th}$, its rate of change quickly drops and appears to saturate at higher condensate densities converging to a value of approximately 0.5. 
\begin{figure}[t!]
	\center	
	\includegraphics[width=1.0\columnwidth]{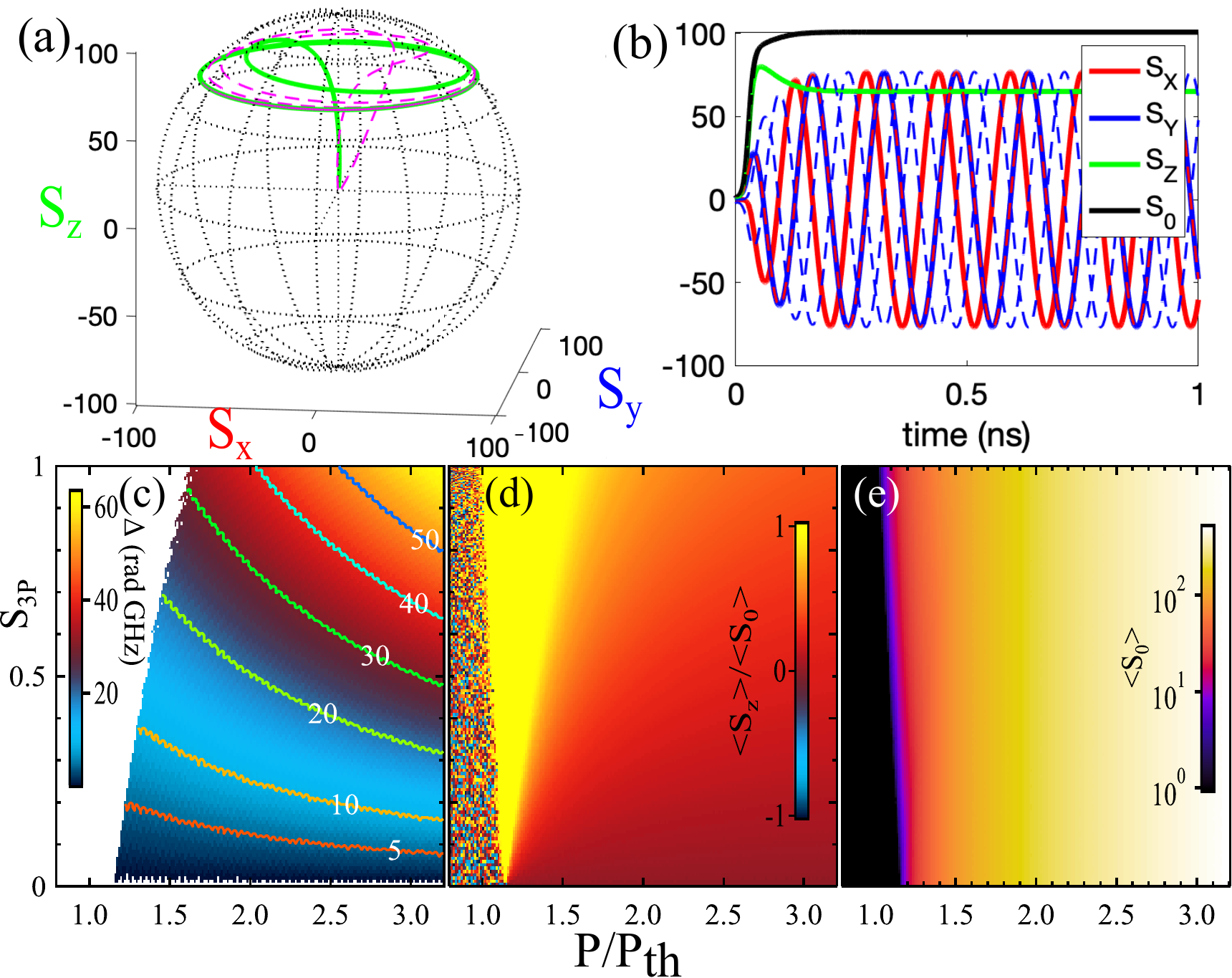} 
	\centering
	\vspace{-20pt}
	\caption{(a) Simulated pseudo-spin trajectories on the Bloch sphere under the self-induced field with different initial starting points (green and purple lines) (a) and temporal evolution of the pseudo-spin stokes components $S_x,S_y,S_z$ and $S_0$ (red, blue, green, and black lines respectively, dotted lines are for different starting order parameter) (b), reproduced from the GPE model. Oscillation frequency (with contour lines) (c), normalized  $\langle S_z \rangle$ stokes component (d) and total density $\langle S_0 \rangle$ (e) for varying power and excitation polarization. $S_{3P}=\sin{2\theta}$ is the $S_3$ component of the excitation polarization and $\theta$ the angle of the QWP, as described in Eq.\eqref{eq.pump}.  }
	\vspace{-15pt}
	\label{fig4}
\end{figure}
We can understand the dynamics of the self-induced Larmor precession by modeling the spinor polariton condensate order parameter through a set of driven-dissipative Gross-Pitaevskii equations coupled to spin-polarized rate equations describing excitonic reservoirs $X_\sigma$ feeding the two spin components of the condensate,
\begin{subequations} \label{eq.orig}
\begin{align} 
i & \dot{\psi}_\sigma  =  \eta_\sigma(t) + \frac{1}{2} \Big[\alpha |\psi_\sigma|^2 + i \left( R X_\sigma - \Gamma \right) \Big] \psi_\sigma \\
& \dot{X}_\sigma  =   - \left( \Gamma_R + R|\psi_\sigma|^2 \right)X_\sigma + \Gamma_s(X_{-\sigma} - X_\sigma) + P_\sigma,
\end{align}
 \end{subequations}
where $\eta_\sigma(t)$ is Gaussian stochastic noise defined by the correlators $\langle d\eta_\sigma(t) d\eta_{\sigma'}(t') \rangle 
=  (\Gamma + R X_\sigma) \delta_{\sigma \sigma'} \delta(t-t')/2$ and $\langle d\eta_\sigma(t) d\eta_{\sigma'}^*(t') \rangle  = 0$. Here, $\alpha$ denotes the same spin polariton-polariton interaction strength, $R$ is the rate of stimulated scattering of polaritons into the condensate, and $\Gamma$ is the polariton decay rate, $\Gamma_R$ and $\Gamma_s$ describe the decay rate and spin relaxation of reservoir excitons~\cite{parameters}. The active reservoir $X_\sigma$, which feeds the condensate, is driven by a background of inactive excitons $P_\sigma$ which do not satisfy energy-momentum conservation rules to scatter into the condensate. Since these inactive excitons also experience spin relaxation $\Gamma_s$ the polarization of $P_\sigma$ will not coincide with that of the incident optical excitation. For the rate equation describing the inactive reservoir dynamics [see SI] one can derive the following steady state expression,
\begin{equation}\label{eq.pump}
\begin{pmatrix} P_+ \\ P_- \end{pmatrix} = \frac{L}{W + 2 \Gamma_s} \begin{pmatrix} W \cos^2{(\theta-\pi/4)} + \Gamma_s \\ W \sin^2{(\theta-\pi/4)} + \Gamma_s \end{pmatrix}
\end{equation}
where $L$ is the power of the optical excitation, $\theta$ is the quarter wave plate (QWP) angle of the experiment determining the polarization of the incident light, and $W$ is a phenomenological spin-conserving redistribution rate of inactive excitons into active excitons ($X_\sigma$). As such, $\theta = \pm 45^\circ$ corresponds to clockwise and anticlockwise circular polarization respectively and $\theta = 0^\circ,90^\circ$ to linear polarization.

We find that the experimental results are due to the self-induced Larmor precession of the condensate pseudospin~\cite{kavokin_quantum_2004},
\begin{equation} \label{eq.S}
\mathbf{S} =\frac{1}{2} \Psi^\dagger  \boldsymbol{\sigma}  \Psi,
\end{equation}
where $\boldsymbol{\sigma} = (\hat{\sigma}_x,\hat{\sigma}_y,\hat{\sigma}_z)$ is the Pauli matrix vector. The components of the pseudospin $\mathbf{S} = (S_x,S_y,S_z)^T$ are written $S_x = \text{Re}{(\psi_-^* \psi_+)}$, $S_y = -\text{Im}{(\psi_-^* \psi_+)}$, and $S_z = (|\psi_+|^2 - |\psi_-|^2)/2$. The total pseudospin is defined $S_0 = \sqrt{S_x^2 + S_y^2 + S_z^2} = (|\psi_+|^2 + |\psi_-|^2)/2$. The DOP and DCP correspond to $\langle S_0 \rangle$ and $\langle S_z \rangle$. The precession appears due to the parity symmetry breaking driving term $P_+ \neq P_-$ which creates a spin imbalanced reservoir and condensate which results in an effective Zeeman splitting of strength $\Omega_z = \alpha S_z$. It is clear that if the pseudo-spin is initially tilted with respect to the $\hat{\mathbf{z}}$ axis of the sample (direction out of the cavity plane) it will start precessing around $\boldsymbol{\Omega} = \Omega_z \hat{\mathbf{z}}$ due to the action of the magnetic field $\hbar \dot{\mathbf{S}} = \boldsymbol{\Omega} \times \mathbf{S}$. 

Results from simulation are shown in~\Fig{fig2}(b,d) showing good agreement with the experiment. The unique feature of the polariton condensate is its driven-dissipative nature which, in hand with the non-linearity $\alpha$, drives the pseudospin always into the same trajectory on the surface of the Bloch sphere for different initial conditions. In other words, the most stable solution of the condensate is a limit cycle whose trajectory is determined by the parameters of the model, including the handles of the experiment such as excitation strength and polarization $\theta$. This is why the same oscillations can be observed in the $g^{(1)}(\tau)$ between different experimental realizations of the condensate (i.e., the precession is strongly reproducible). In Fig.~\ref{fig3}(d) we plot the extracted period and amplitude of the simulated precession for $\eta_\sigma = 0$ showing a qualitative match with experimental observables. 

We plot the trajectory of the condensate pseudo spin on a Bloch sphere, \Fig{fig4}(a), as well as individual components, in \Fig{fig4}(b). As the condensate density builds-up, the pseudospin of the system relaxes to a fixed periodic trajectory on the Bloch sphere within $\approxeq 200$ ps, under the influence of the self induced field. The final stable limit cycle is not affected by the spontaneous symmetry breaking during condensation, that results in different starting points for the trajectory of the system order parameter (\Fig{fig4}(a) green and purple lines) in individual condensation realizations. This suggests that the precession can not be resolved through time synchronized experiments, integrating over different realizations, as every time the oscillation can have an arbitrary starting point (see \Fig{fig4}(b) where solid and dashed lines depict the different realizations). The simulation also reveals that the pseudo-spin trajectory will either converge to a fixed point attractor on the Bloch sphere with no oscillations (overdamped), reach a stable limit cycle (stable precession), or will undergo a damped precession eventually converging to a fixed point attractor (underdamped) [See SI]. Such regimes can be differentiated by detection of the condensate DOP~\cite{gnusov_optical_2020}. In Fig.~\ref{fig4}(c) we show the difference in energy $\Delta$ between the $\psi_\pm$ polaritons (i.e., the difference between their strongest spectral peaks) extracted directly from simulation of the dynamics. The reason we investigate the dynamics to obtain $\Delta$ is because $\alpha S_z \neq 0$ does not imply a pseudo-spin precession [e.g., see Fig.~\ref{fig2}(a)]. In Fig.~\ref{fig4}(d) and \ref{fig4}(e) we show the time averaged pseudospin components $\langle S_z \rangle / \langle S_0 \rangle$ and $\langle S_0 \rangle$ respectively, where the averaging is done over a 50 ns time window. The speckled region at low power in Fig.~\ref{fig4}(d) is due to the condensate being below threshold resulting in a stochastic normalized pseudo-spin value.

We point out that the limit cycle state corresponding to $\Delta \neq 0$ and $\langle S_{x,y}\rangle = 0$ can be expressed analytically by considering the ansatz,
\begin{equation}
\ \Psi = \begin{pmatrix} \psi_+ e^{-i\omega_+ t} \\ \psi_- e^{-i \omega_- t} \end{pmatrix},\
\end{equation}
where $\omega_\sigma \in \mathbb{R}$, $\dot{\psi_\sigma} = 0$, and $\dot{X}_\sigma = 0$. This is a limit cycle solution to Eq.~\eqref{eq.orig} can be expressed in terms of parameters of the model,
\begin{equation}
\omega_\sigma    =  \frac{\alpha}{2} |\psi_\sigma|^2, \quad X_\sigma = \frac{\Gamma}{R}, \quad  |\psi_\sigma|^2  = \frac{P_\sigma}{\Gamma} - \frac{\Gamma_R  }{R}.
\end{equation}
The solution of the condensate corresponds to cancellation between gain and dissipation on a specific latitude line on the Bloch sphere of an angle $\varphi = \sin^{-1}{(S_z/S_0)} =  \sin^{-1}{[(P_+ - P_-)/(P_+ + P_- - 2 \Gamma \Gamma_R/R)]}$. A continuous real spectrum of $\omega_\sigma$ therefore belongs to these degenerate latitude lines in our driven-dissipative condensate which can be tuned through the excitation polarization. Visibility oscillation amplitude can be shown to be $1-\sin(\varphi)$, and the oscillation period  is given by $2\pi/(\omega_+-\omega_-)$ (see Fig.~\ref{fig3}(d)).

In conclusion, we have revealed the collapse and revival of coherence in a polariton condensate. The amplitude and period of these revivals are dynamically tunable with the total interactions in the condensate. We attribute this effect to self induced Larmor precession due to polariton interactions and find that the system undergoes more than $10^5$ oscillations with surprising stability, despite the effects of gain and dissipation and are fully reproduced with coupled condensate-reservoir mean field theory. It would be interesting to study the temporal-mixing of quasi-degenerate (with same principal quantum number $n$) trapped states that have been proposed, but not experimentally observed, as a pathway towards simulating a flux qubit with polariton condensates~\cite{xue_split-ring_2020, sedov_persistent_2020}. Our observations, reveal for the first time that this gain dissipative system in dynamic equilibrium, can display persistent coherent oscillations in the pseudo-spin Bloch sphere for the duration of the quasi-CW optical excitation pulse, in-spite of ultra-fast single particle lifetimes, and opens new interesting applications for polariton condensates such as optical magnetometry~\cite{budker_optical_2007}. This can be achieved using magnetic microcavities with strong exciton $g$-factor~\cite{rousset_magnetic_2017} and by utilizing electrical tuning of the exciton mode and polariton non-linearity~\cite{christmann_oriented_2011, tsintzos_electrical_2018}. Moreover, it brings to the forefront intriguing possibilities, such as the creation of spin-squeezed states, in a similar fashion to what has been accomplished already with atomic condensates~\cite{laudat_spontaneous_2018}, and whether a truly PT-symmetric state~\cite{chestnov_permanent_2016, kalozoumis_coherent_2020} can be engineered in this non-hermitian platform. 

\section*{Acknowledgements}
The authors declare no competing financial interests. A.A. would like to acknowledge useful discussions with N. Berloff, S. Tsintzos., J. T{\"o}pfer and S. Baryshev. N.A.G.  acknowledges support of RFBR project 18-29-20032. P.G.L acknowledges the support of the UK’s Engineering and Physical Sciences Research Council (grant EP/M025330/1 on Hybrid Polaritonics). 


%
%

\end{document}